\documentclass[12pt]{amsart}
\usepackage{amsfonts,amssymb,amsmath}
\usepackage[foot]{amsaddr}
\usepackage{epsfig}

\newcommand\RR{\mathbb R}
\newcommand\CC{\mathbb C}
\renewcommand{\Re}{\mathop{\mathrm{Re}}}
\renewcommand{\Im}{\mathop{\mathrm{Im}}}

\newcommand\beq{\begin{equation}}
\newcommand\eeq{\end{equation}}
\newcommand{\ba}{\begin{array}}
\newcommand{\ea}{\end{array}}
\newcommand{\bea}{\begin{eqnarray}}
\newcommand{\eea}{\end{eqnarray}}

\newtheorem{theorem}{Theorem}
\newtheorem{lemma}{Lemma}
\newtheorem{prop}{Proposition}

\begin{document}

\title[Holomorphic eigenfunctions]{Holomorphic eigenfunctions of the vector field associated with 
the dispersionless Kadomtsev-Petviashvili equation}
\thanks{The first author was partially supported by the  the Russian Federation Government grant 
No~2010-220-01-077, RFBR grant 11-01-12067-ofi-m-2011, by the 
program  ``Fundamental problems of nonlinear dynamics'' of the Presidium of RAS and by INFN}
\author{P. G. Grinevich$^1$ } 
\address{1.~L.D.Landau Institute for Theoretical Physics RAS}  
\author{P. M. Santini$^2$} 
\address{2.~Dipartimento di Fisica, Universit\`{a} di Roma "La Sapienza" and
Istituto Nazionale di Fisica Nucleare, Sezione di Roma}
\maketitle

\begin{abstract}
Vector fields naturally arise in many branches of mathematics and physics. Recently it was 
discovered that Lax pairs for many important multidimensional integrable partial differential 
equations  (PDEs) of hydrodynamic type (also known as dispersionless PDEs) consist of vector 
field equations. These vector fields have complex coefficients and their analytic, in the 
spectral parameter, eigenfunctions play an important role in the formulations of the direct and 
inverse spectral transforms.

In this paper we prove existence of eigenfunctions of the basic vector field associated with 
the celebrated dispersionless Kadomtsev-Petviashvili equation, which are holomorphic in the spectral 
parameter $\lambda$ in the strips 
$|\Im\lambda|> C_0$. 
\end{abstract}

\section{Introduction}

The dispersionless Kadomtsev-Petviashvili equation (dKP) 
\beq
\label{dKP}
(u_t+uu_x)_x+u_{yy}=0,~~~
u=u(x,y,t)\in\RR,~~~~~x,y,t\in\RR,
\eeq 
is a model equation describing the propagation of weakly nonlinear quasi one-dimensional waves, 
in the absence of dispersion and dissipation, in many physical contexts 
(see \cite{Timman}, \cite{KZ}, \cite{KP}, \cite{AC}). It arises as commutation condition 
for the following pair of vector fields:
\beq
\label{L1L2}
\ba{l}
\hat L_1\equiv \partial_y+\lambda\partial_x-u_x\partial_{\lambda}, \\
\hat L_2\equiv \partial_t+(\lambda^2+u)\partial_x+(-\lambda u_x+u_y)\partial_{\lambda},
\ea
\eeq
$\lambda\in\CC$ being the spectral parameter \cite{KG}, \cite{Kri}, \cite{Zak}. 
This integrability scheme allows one to construct the formal solution of the Cauchy problem for dKP, 
and properly defined analytic in $\lambda$ eigenfunctions of the above vector fields play a 
crucial role in the formulation of the direct and inverse spectral transforms \cite{Manakov3}, \cite{Manakov6}.

In this paper, in the framework of the direct problem, we construct analytic zero-level eigenfunctions 
of the first vector field $\hat L_1$ (therefore we omit the $t$-dependence in all formulas):
\beq
\label{eq:vf1}
\hat L_1 \Psi=[\partial_y+\lambda\partial_x-u_x(x,y)\partial_{\lambda}]\Psi = 0,
\eeq
where 
\beq
\Psi=\Psi(x,y,\lambda), \ \ x,y\in\RR, \ \ \lambda\in\CC,
\eeq
with the asymptotic behaviors
\beq
\label{eq:asympt1.1}
\Psi_1(x,y,\lambda)\rightarrow\lambda, \ \ \mbox{as} 
\ \ x^2+y^2\rightarrow\infty,
\eeq
\beq
\label{eq:asympt1.2}
\Psi_2(x,y,\lambda)\rightarrow x-\lambda y, \ \ \mbox{as} 
\ \ x^2+y^2\rightarrow\infty,
\eeq
assuming that $u(x,y)$ decays sufficiently fast if 
$x^2+y^2\rightarrow\infty$. A basis of common eigenfunctions for both vector fields $\hat L_1$ 
and $\hat L_2$ is given by  
$\Psi_1$ and by the combination $\Psi_2-t\Psi_1^2$ \cite{Manakov3}. 

To construct these analytic eigenfunctions, we make use of the complex forced Hopf equation 
(\ref{eq:ham8}), describing  the level sets of function $\Psi$. We remark that equation 
(\ref{eq:vf1}) is equivalent to the Benney system \cite{B} written in terms of the generating function 
for the momenta, and in the Benney framework, $\Psi-\lambda$ is automatically holomorphic 
near infinity \cite{KM1}. The relation between (\ref{eq:vf1}) and the complex version of 
(\ref{eq:ham8}) was also introduced and used in the framework of Benney system \cite{KM1}; in this 
contest the solution of  (\ref{eq:ham8}) is also automatically holomorphic near infinity. The vector 
field $\hat L_2$ is equivalent to the symmetry of the Benney system, introduced in \cite{KM2}.

The procedure we propose here to 
construct analytic eigenfunctions of vector field equations should be applicable also to the 
vector field Lax pairs of other basic 
examples of integrable PDEs, 
like the heavenly \cite{Pleb}, the two dimensional dispersionless Toda \cite{FP,Zak1,BF} 
and the Martinez Alonso - Shabat - Pavlov \cite{MA-S,Pavlov} equations.  

We remark that Derchyi Wu has recently proven the unique solvability of the 
nonlinear Riemann-Hilbert problem characterizing the inverse transform for dKP, under a 
small-norm assumption, using appropriate Sobolev spaces  (private communication).   

We are very grateful to S.V.Manakov for having constantly stimulated this research, and for
many valuable discussions.

\section{Three approaches to real vector field equations}

To start with, let us recall some basic facts from Hamiltonian mechanics.
In the  {\bf real} framework, $\hat L_1$ is a Hamiltonian vector field
\beq
\label{eq:ham1}
\hat L_1\equiv \partial_y+\lambda\partial_x-u_x(x,y)\partial_{\lambda}=\partial_y+
\{H,\cdot \}_{\lambda,x}, \ \ x,y,\lambda\in\RR, \ \ u\in\RR,
\eeq
where 
\beq
\label{eq:ham2}
H(x,\lambda,y)=\frac{\lambda^2}{2}+u(x,y)
\eeq
is the Hamiltonian of a newtonian particle in the time-dependent potential $u(x,y)$ 
(here time is $y$). 
Eigenfunctions $\Psi$ of $\hat L_1$, 
\beq
\label{eq:ham2.1}
\hat L_1\Psi=0,  \ \ x,y,\lambda\in\RR, \ \ \Psi(x,y,\lambda)\in\RR,
\eeq
are exactly conservation laws for the associated hamiltonian dynamical system:
\beq
\label{eq:ham3}
\frac{d x}{d y}=\lambda, \ \ \frac{d\lambda}{d y}=-u_x(x,y).
\eeq
Hamiltonian systems can also be studied using the Hamilton-Jacobi equation
\beq
\label{eq:ham4}
\frac{\partial S}{\partial y}+H\left(x,\frac{\partial S}{\partial x}\right)=0, \ \ \lambda= \frac{\partial S}{\partial x},
\eeq
which, in our example, takes this form:
\beq
\label{eq:ham5}
\frac{\partial S}{\partial y}+\frac{1}{2}\left(\frac{\partial S}{\partial x}\right)^2  + u(x,y)=0.
\eeq
Equations (\ref{eq:ham2.1}), (\ref{eq:ham3}), (\ref{eq:ham5}) give three equivalent formulations 
of the problem. If we know the trajectories of (\ref{eq:ham3}), then any function constant on these 
trajectories solves (\ref{eq:ham2.1}). And vice versa, if we have two independent constants 
of motion, the common level sets are the trajectories. The connection between the dynamical 
system  (\ref{eq:ham3}) and the Hamilton-Jacobi equation (\ref{eq:ham5})  is well-known. 
Finally, we go from (\ref{eq:ham2.1}) to (\ref{eq:ham5}) just considering the level sets 
\beq
\label{eq:ham6}
\Psi(x,y,\lambda)=k;
\eeq
solving (\ref{eq:ham6}) with respect to $\lambda$:
\beq
\label{eq:ham7}
\lambda=\Lambda(x,y,k)
\eeq
we construct a function $\Lambda$ satisfying the forced Hopf equation
\beq
\label{eq:ham8}
\frac{\partial \Lambda}{\partial y}+\Lambda\frac{\partial\Lambda}{\partial x}=-\frac{\partial u}{\partial x},
\eeq
which is nothing but the $x$-derivative of the Hamilton-Jacobi equation  (\ref{eq:ham5}) for $\Lambda=S_x$. 

Using the equivalence between (\ref{eq:ham2.1}) and  (\ref{eq:ham3}), we can easily construct 
Jost eigenfunctions of (\ref{eq:vf1}) (another important ingredient of the dKP direct problem 
\cite{Manakov3}) satisfying the following asymptotics 
\beq
\label{eq:ham9}
\begin{aligned}& \varphi_1(x,y,\lambda) \rightarrow\lambda, \\
& \varphi_2(x,y,\lambda) \rightarrow x-\lambda y,
\end{aligned}
\ \ \  \ \ \mbox{as} \ \ y\rightarrow-\infty.
\eeq
Indeed, consider the solution of (\ref{eq:ham3}) with the following initial data
\beq
\label{eq:ham10}
x=x_0, \ \ \lambda=\lambda_0 \ \  \mbox{at}  \ \ y=y_0.
\eeq
For sufficiently regular and well-localized $u(x,y)$, the solution exists globally in $y\in\RR$, is unique 
and has the free particle behavior at $\pm\infty$
\beq
\label{eq:ham11}
\lambda=\lambda_\pm(x_0,\lambda_0,y_0)+o(1), x=x_\pm(x_0.\lambda_0,y_0)+y\lambda_\pm(x_0,\lambda_0,y_0)+ o(1)  
\ \mbox{as}  \  y\rightarrow\pm\infty.
\eeq
Therefore the Jost solutions can be directly expressed in terms of the long-time asymptotics:
\beq
\label{eq:ham12}
\varphi_1(x_0,y_0,\lambda_0)= \lambda_-(x_0,\lambda_0,y_0), \ \
\varphi_2(x_0,y_0,\lambda_0)= x_-(x_0.\lambda_0,y_0).
\eeq
\section{Reduction to the complex Hopf equation}  

If $\lambda$ is complex, we loose one of the principal ingredients -- the dynamical system (\ref{eq:ham3}).
One of the reasons is that we do not want to assume that $u(x,y)$ has good analytic continuation 
to the complex domain. Nevertheless, the connection between equations  (\ref{eq:vf1}) and (\ref{eq:ham8})
is still present and it will be our principal tool.
\begin{lemma}
At regular points ($\partial_{\lambda}\Psi(x,y,\lambda)\ne0$) the function $\Lambda(x,y,k)$ defined by 
(\ref{eq:ham6}), (\ref{eq:ham7}) for complex $\lambda$, satisfies the {\bf complex} Hopf 
equation with source:
\beq
\label{eq:hopf1}
\Lambda_y+\Lambda\Lambda_x=-u_x.
\eeq
\end{lemma}
The proof is straightforward:
$$
\Psi(x,y,\Lambda(x,y,k))\equiv k
$$
therefore
$$
\frac{d}{dx}\Psi(x,y,\Lambda(x,y,k))=0, \ \ 
\frac{d}{dy}\Psi(x,y,\Lambda(x,y,k))=0, 
$$
$$
\Psi_x+\Lambda_x\Psi_{\lambda}=0, \ \ \Psi_y+\Lambda_y\Psi_{\lambda}=0,
$$
$$
\Lambda\Psi_x+\Lambda\Lambda_x\Psi_{\lambda}+\Psi_y+\Lambda_y\Psi_{\lambda}=0,
$$
$$
\Lambda\Psi_x+\Lambda\Lambda_x\Psi_{\lambda}+\Psi_y+\Lambda_y\Psi_{\lambda}-
u_x\Psi_{\lambda}+u_x\Psi_{\lambda} =0,
$$
$$
[\Psi_y+\Lambda\Psi_x-u_x\Psi_{\lambda}]+
[\Lambda_y\Psi_{\lambda}+\Lambda\Lambda_x\Psi_{\lambda}+u_x\Psi_{\lambda}] =0,
$$
$$
[\Lambda_y+\Lambda\Lambda_x+u_x]\Psi_{\lambda} =0.
$$
Taking into account that $\Psi_{\lambda}\ne0$,  we have the Hopf equation. \qed

Consider the level sets for the function $\Psi_1(x,y,\lambda)$; the
asymptotic condition (\ref{eq:asympt1.1}) implies:
$$
\lambda+o(1) =k , \ \ \mbox{as} 
\ \ x^2+y^2\rightarrow\infty;
$$
therefore
\beq
\label{eq:asympt2.1}
\Lambda_1(x,y,k)\rightarrow k \mbox{\ \  as \ \ } x^2+y^2\rightarrow\infty.
\eeq
Therefore, to construct  $\Psi_1$, we solve the Hopf equation 
\beq
\label{eq:hopf2}
(\Lambda_1)_y+\Lambda_1(\Lambda_1)_x=-u_x
\eeq
with the boundary condition:
\beq
\label{eq:asympt3}
\Lambda_1(x,y,k)\rightarrow k \mbox{\ \ \  as \ \ } x^2+y^2\rightarrow\infty.
\eeq

Unfortunately we can not apply the same procedure to the function $\Psi_2$ 
because, for $u\equiv 0$, $\Psi_2(x,y,\lambda)\equiv x-\lambda y$ and the levels 
sets 
$$
x-\lambda y = k, \ \ \lambda = \frac{x-k}{y}
$$
are always singular. 

Instead, let us consider the level sets for the function
$$
\Psi(x,y,\lambda)=\Psi_1(x,y,\lambda)+ \varepsilon \Psi_2(x,y,\lambda) , \ \ 
\varepsilon \ll 1,
$$
$$
\Psi(x,y,\Lambda_1+\varepsilon\Lambda_2+O (\varepsilon^2))=k,
$$
where $k=\Psi_1(x,y,\Lambda_1)$.
Combining the coefficients at order $\varepsilon^1$ we obtain:
\beq
\label{eq:psi2def}
\Psi_2(x,y,\Lambda_1(x,y,k))=-\Lambda_2(x,y,k)\left.\partial_{\lambda}
\Psi_1(x,y,\lambda)\right|_{\lambda=\Lambda_1(x,y,k)}.
\eeq
For $x^2+y^2\rightarrow\infty$ we obtain
$$
\Lambda_2(x,y,k)\rightarrow -\Psi_2(x,y,\Lambda_1(x,y,k)) 
\mbox{\ \  as \ \ } x^2+y^2\rightarrow\infty;
$$
therefore
\beq
\label{eq:asympt2.2}
\Lambda_2(x,y,k)\rightarrow -(x-ky) \mbox{\ \  as \ \ } x^2+y^2\rightarrow\infty.
\eeq
Substituting $\Lambda=\Lambda_1+\varepsilon\Lambda_2+O(\varepsilon^2) $ into the Hopf equation 
(\ref{eq:hopf1}) and combining the coefficients at order $\varepsilon^1$ we obtain:
\beq
\label{eq:hopf2.2}
(\Lambda_2)_y+(\Lambda_1\Lambda_2)_x=0.
\eeq

Therefore, to construct $\Psi_2(x,y,\lambda)$, we have to solve the linearized Hopf 
equation (\ref{eq:hopf2.2}) with the boundary condition (\ref{eq:asympt2.2}).

\section{A sketch of our construction}

Here we present a short description of our work.

\begin{enumerate}
\item Assuming that $|\Im k|$ is sufficiently large, we solve the nonlinear Hopf equation 
(\ref{eq:hopf2}), (\ref{eq:asympt3}), and obtain some estimates on the solution. 

To solve equation (\ref{eq:hopf2}), and, in general, nonlinear PDEs, it is natural to use Sobolev spaces 
not only because one can control the derivatives involved in the equations, but also
because Sobolev spaces with sufficiently many derivatives are Banach algebras -- the product of 
two elements of that space also belongs to it, and the multiplication is continuous. More precisely, 
if $f(x,y),g(x,y)\in H^{l}(\RR^2)$, $l\ge 2$, then 
$f\cdot g\in H^{l}(\RR^2)$ and 
$\|f\cdot g\|_{H^{l}(\RR^2)}\le\alpha_l\|f\|_{H^{l}(\RR^2)} \|g\|_{H^{l}(\RR^2)}$
(see Lemma~\ref{lem:l7}  of the  Appendix for details).

Assuming that $u(x,y)\in H^{4}(\RR^2)$, we construct the solution of (\ref{eq:hopf2}), (\ref{eq:asympt3})
in the form $\Lambda_1(x,y,k)=k+\phi(x,y,k)$, where $\phi\in H^{4}(\RR^2)$.

\item To obtain some preliminary estimates on the behavior of $\phi(x,y,k)$ at large $x^2+y^2$, we show 
that  $\phi(x,y,k)$ lies also in the Banach spaces $W^{2,p}$ for $p$ sufficiently close to 2. 

\item Assuming that $u(x,y)$ decays sufficiently fast at infinity (see Proposition~\ref{prop:estphi} 
for details), we show  that $\phi(x,y,k)$  satisfies an inhomogeneous Beltrami equation. 
Using this fact and the estimates on $\phi(x,y,k)$  proven in the previous step, we obtain 
additional estimates on the asymptotics of $\phi(x,y,k)$ for large $x^2+y^2$. 

\item By interpreting the linearized Hopf equation (\ref{eq:hopf2.2}) as an inhomogeneous Beltrami equation, 
we prove the existence and uniqueness of $\Lambda_2(x,y,k)$. 

\item Using the linearized Hopf equations for the functions $\partial_{\bar k}\Lambda_1$,  $\partial_{\bar k}\Lambda_2$, 
we show that $\partial_{\bar k}\Lambda_1=\partial_{\bar k}\Lambda_2=0$; i.e., for fixed $(x,y)$,  the functions
$\Lambda_1$,  $\Lambda_2$ are analytic in $k$.

\item Using the linearized Hopf equations for the function $\partial_{k}\Lambda_1$, we show that, for sufficiently
large $|\Im k|$, we have  $|\partial_{k}\Lambda_1-1|\le C<1$. Combining this fact with estimates on $\phi(x,y,k)$,
we show that the inversion with respect to $k$ of equation (\ref{eq:ham7}) is well-defined, and gives us 
the analytic eigenfunctions $\Psi_1(x,y,\lambda)$, $\Psi_2(x,y,\lambda)$ of the vector field $\hat L_1$.
\end{enumerate}

\section{Notation, some basic results and definitions}

Assuming that $k$ is a fixed complex number, $\Im k\ne 0$, we shall use the following complex notation:
\begin{align}
\label{eq:complex1}
&z=  \frac{1}{\bar k - k }(x-ky), 
&\bar{z}           &=   -\frac{1}{\bar k - k }(x-\bar k y), \nonumber \\
&x= \bar k z + k \bar z, 
&y                 &= z + \bar z, \nonumber \\
&\partial_{\bar z}= \partial_y+k\partial_x,  
&\partial_z        &=  \partial_y+\bar k\partial_x, \nonumber\\ 
&\partial_x= \frac{1}{\bar k -k}(\partial_{z}-\partial_{\bar z}), 
&\partial_y        &=\frac{1}{\bar k -k}(\bar k \partial_{\bar z}- k\partial_{z}).  \nonumber \\ 
\end{align}
To simplify the notation, we shall use the following agreement: unless it generates confusion,
$f(x,y)$ and $f(z,\bar z)$ will denote {\bf the same} function in the plane. Moreover we will 
often omit the $\bar z$ dependence in the argument of functions; therefore writing $f(z)$ instead of
$f(z,\bar z)$ {\bf does not imply} $\partial_{\bar z} f(z)=0$. 

We use the following normalization for the Fourier transform:
\beq
\label{eq:fourier1}
f(x,y) =\frac{1}{2\pi}\iint \hat f(p) e^{i(p_x x+ p_y y)} 
d^2 p, \ \ p=(p_x,p_y),  \ \ d^2p = d p_x d p_y,
\eeq
\beq
\label{eq:fourier2}
\hat f(p) =\frac{1}{2\pi}\iint f(x,y) e^{-i(p_x x+ p_y y)} dx dy.
\eeq

It is clear, that
\beq
\label{eq:fourier3}
\widehat{\partial_z f} = i(p_y+\bar k p_x) \hat f, \ \ 
\widehat{\partial_{\bar z} f} =  i(p_y+ k p_x)  \hat f,
\eeq
\beq
\label{eq:fourier4}
\widehat{fg}(p) = \frac{1}{2\pi} (\hat f* \hat g)(p),
\eeq
where
\beq
\label{eq:fourier4.1}
(\hat f*\hat g)(p)=\iint \hat f(p-q) \hat g(q) d^2 q
\eeq
is the {\bf convolution} operator.

We recall that, if $\hat f\in L^1(\RR^2)$, then $f(z)$ is continuous, decays 
for $|z|\rightarrow\infty$ and
\beq
\label{eq:fourier5}
|f(z)|\le \frac{1}{2\pi}\iint |\hat f(p)| d^2p.
\eeq

We also need the following theorem from the Vekua's book \cite{Vekua}. 
\begin{theorem}
\label{th:vekua1}
Denote by $\Pi$ the following operator: $\Pi=\partial_{z}\partial^{-1}_{\bar z}$
\beq
\label{eq:belt1}
(\Pi f)(z)= -\frac{1}{\pi}\iint \frac{f(\xi)}{(\xi-z)^2} d^2\xi
\eeq
\begin{enumerate}
\item Let $f(\xi)\in L^p(\RR^2)$. Then the integral (\ref{eq:belt1}) is well-defined, in the sense
of principal value, almost everywhere in $z$ and $(\Pi f)(z) \in L^p(\RR^2)$.

\item For all $1<p<\infty$, $\Pi$ is a bounded operator on $L^p(\RR^2)$
\beq
\label{eq:belt2}
\|\Pi f\|_{L^p}\le \gamma(p) \|f\|_{L^p},
\eeq
where $\gamma(p)$ is a continuous function of $p$ and $\gamma(2)=1$.
\end{enumerate}
\end{theorem}

The proof of this Theorem essentially uses the results of the papers \cite{Riesz}, \cite{Titchmarsh}. This statement can also be viewed as a corollary of the Zygmund-Calderon theorem 
(see \cite{Stein}).

The Sobolev space $H^l(\RR^2)=W^{l,2}(\RR^2)$ is the Hilbert space defined using the 
following scalar product
\beq
\label{eq:sobolev1}
(f,g)_{H^l}=\iint \overline{\hat f(p)} \hat g(p) (1+p_x^2+p_y^2)^l d^2 p
\eeq
\beq
\label{eq:sobolev2}
\|f\|_{H^l}=\sqrt{(f,f)_{H^l}}
\eeq

Another (equivalent) norm is defined by 
\beq
\label{eq:sobolev2.1}
\|f\|_{W^{l,2}} =\left[
\sum\limits_{k_1+k_2\le l} \iint |\partial_{x}^{k_1} \partial_{y}^{k_2} f(x,y)|^2 dx dy \right]^{1/2}
\eeq

The Sobolev space $W^{l,p}(\RR^2)$ is a Banach space generated by the following norm:
\beq
\label{eq:sobolev2.3}
\|f\|_{W^{l,p}} =\left[
\sum\limits_{k_1+k_2\le l} \iint |\partial_{x}^{k_1} \partial_{y}^{k_2} f(x,y)|^p dx dy \right]^{1/p}
\eeq

We shall also use the following notations $f\in W^{l,2\pm\epsilon_0}$, $f\in W^{l,2\pm}$:  
\begin{enumerate}
\item $f\in W^{l,2\pm\epsilon_0}$  if $f\in W^{l,2+\epsilon}$ for all  $\epsilon$ such 
that $-\epsilon_0\le\epsilon\le\epsilon_0$.
\item $f\in W^{l,2\pm}$ if $f\in W^{l,2\pm\epsilon_0}$ for some $\epsilon_0>0$.
\end{enumerate}

\section{Solving the Hopf equation for $\Lambda_1$}

Let us describe the iterative procedure for solving
(\ref{eq:hopf2}). Let 
\beq
\Lambda_1=k + \phi(x,y,k);
\eeq
then we have:
\beq
\label{eq:hopf3}
\phi_y+k\phi_x +\phi\phi_x=-u_x.
\eeq

In the complex notation (\ref {eq:complex1}),   equation (\ref{eq:hopf3}) 
takes the form:
\beq
\label{eq:hopf4}
\partial_{\bar z} \phi +\frac{1}{\bar k -k}   \phi (\partial_z-\partial_{\bar z})\phi=-\frac{1}{\bar k -k}  
(\partial_z-\partial_{\bar z}) u,
\eeq
equivalent, for $\phi$ decaying at infinity, to
\beq
\label{eq:hopf5}
\phi = - \frac{1}{2(\bar k -k)} ( \partial_{\bar z}^{-1}\partial_{z} -1) (\phi^2)  - 
\frac{1}{\bar k -k} ( \partial_{\bar z}^{-1}\partial_{z}-1) (u).
\eeq

Let us check that, for sufficiently small $\phi$, the map
\beq
\label{eq:hopf6}
\phi \rightarrow G(\phi) = \frac{1}{2(\bar k -k)} ( \partial_{\bar z}^{-1}\partial_{z} -1) (\phi^2)
\eeq
in the space $H^l(\RR^2)$ is contracting.

In the Fourier representation we have:

\beq
\label{eq:invbar1}
\widehat{\partial_{\bar z}^{-1}  \partial_{z} f} = \frac{p_y+\bar k p_x}{p_y+k p_x } \hat f,
\eeq
therefore $\partial_{\bar z}^{-1}  \partial_{z}$ is an unitary operator on $H^l(\RR^2)$. Therefore we have
\beq
\label{eq:invbar2}
\left\| \frac{1}{2(\bar k -k)} ( \partial_{\bar z}^{-1}\partial_{z} -1)\right\|_{H^l}\le 
\frac{1}{2|\Im k|}.
\eeq
Consequently, from 
\beq
\label{eq:invbar3}
G(\phi_1)-G(\phi_2) = \frac{1}{2(\bar k -k)} ( \partial_{\bar z}^{-1}\partial_{z} -1)
[(\phi_1+\phi_2)  (\phi_1-\phi_2)] 
\eeq
it follows that (see Lemma~\ref{lem:l7}, part 2) of the Appendix) 
\beq
\label{eq:invbar4}
\|G(\phi_1)-G(\phi_2)\| \le \frac{\alpha_l}{|\Im k|}
(\|\phi_1\|+\|\phi_2\|) \|\phi_1-\phi_2\|.
\eeq
Therefore we proved
\begin{lemma}
Let ${\mathcal B}$ denote the ball in  $H^l$ of radius  $\frac{|\Im k|}{\alpha_l}$:
\beq
\label{eq:invbar5.1}
\phi\in{\mathcal B} \mbox{\ \ if \ \ }   \|\phi\|_{H^l} < \frac{|\Im k|}{\alpha_l}. 
\eeq
\begin{enumerate}
\item
Let $\phi$ be a function from ${\mathcal B}$. Then 
\beq
\label{eq:invbar6}
\|G(\phi) \|_{H^l} \le \frac{1}{2}\|\phi\|_{H^l} 
\eeq
\item
Let $\phi_1$,  $\phi_2$ be functions from the interior of ${\mathcal B}$. Then 
\beq
\label{eq:invbar6}
\|G(\phi_1)-G(\phi_2)\|_{H^l} < \|\phi_1-\phi_2\|_{H^l} 
\eeq
\end{enumerate}
\end{lemma}

\begin{lemma}

\begin{enumerate}
\item 
Let
\beq
\label{eq:invbar7}
\|u\|_{H^l} < C \frac{|\Im k|^2}{2\alpha_l}, \ \ C<1.
\eeq
Then the map
\beq
\label{eq:invbar8}
F(\phi)= - \frac{1}{2(\bar k -k)} ( \partial_{\bar z}^{-1}\partial_{z} -1) (\phi^2)  - 
\frac{1}{\bar k -k} ( \partial_{\bar z}^{-1}\partial_{z}-1) (u)     
\eeq
maps the ball ${\mathcal B}$ onto itself.
\item Let us introduce also the ball ${\mathcal B}_1$ in $H^l$ of radius $\frac{2}{|\Im k|}\|u\|$, which is 
much smaller than ${\mathcal B}$ for large $\Im k$. Then, if 
\beq
\label{eq:invbar8.1}
\|\phi\|_{H^l} < \frac{2}{|\Im k|}\|u\|_{H^l},
\eeq
it follows that $F(\phi)$ maps the ball  ${\mathcal B}_1$ onto itself:
\beq
\label{eq:invbar8.2}
\|F(\phi)\|_{H^l} < \frac{2}{|\Im k|}\|u\|_{H^l}.
\eeq
\end{enumerate}
\end{lemma}
Combining all these estimates we obtain
\begin{theorem}
Let function $u(x,y)$ satisfy the inequality (\ref{eq:invbar7}). Then the iteration procedure
\begin{enumerate}
\item $\phi_0=0$
\item $\phi_{j+1}= F(\phi_j), \ \ j\ge 0$.
\end{enumerate}
converges in $H^l(\RR^2)$, defining the unique localized solution of (\ref{eq:hopf4}), and
\beq
\label{eq:invbar8.3}
\|\phi\|_{H^l} < C\frac{|\Im k|}{\alpha_l}, \ \ \max\limits_{z\in\CC} |\phi(z,\bar z)|< C|\Im k|.
\eeq
\end{theorem}

For the second estimate of (\ref{eq:invbar8.3}) see the Lemma~\ref{lem:l7}, parts 1), 2) of the Appendix.

We finally observe that
\beq
\label{eq:invbar9}
\partial_{\bar z}^{-1} = \frac{1}{k} \left(\partial_x+ \frac{1}{k} \partial_y \right)^{-1},
\eeq
and we can formally write, in the large $\Im k$ limit, 
\beq
\label{eq:invbar10}
\partial_{\bar z}^{-1} = \frac{1}{k}\partial_x^{-1}    \left(\sum \limits_{n\ge 0}
(-1)^n \left(\frac{\partial_y\partial_x^{-1}}{k}\right)^n \right).
\eeq
Then equation (\ref{eq:hopf5}) implies the following formal asymptotics
\beq
\label{eq:invbar11}
\Lambda_1=k-\frac{u}{k}+\frac{\partial_y\partial_x^{-1} u}{k^2}-
\frac{\partial_y^2\partial_x^{-2} u + u^2/2   }{k^3} 
+O\left(\frac{1}{k^4} \right).
\eeq

\section{Some estimates on the asymptotics of $\phi$}
Since, as it will be shown in Section~\ref{sec:s8}, the linear equation for $\Lambda_2$ 
can be interpreted as a Beltrami equation, it follows that Hilbert-Sobolev spaces are not 
adequate to deal with the problem (see \cite{Vekua}). Therefore in this Section we show that 
$\phi$ belongs also to some non-Hilbert Sobolev spaces.

In this section we assume that $l=4$, and the inequality (\ref{eq:invbar7}) is fulfilled.
It means that $\phi\in H^4(\RR^2)$ and we have the inequalities (\ref{eq:invbar8.3}). 
In addition, $\phi$ and its first two derivatives are continuous and bounded 
(see Lemma~\ref{lem:l7}, part 1) of the Appendix).

Let us prove the following estimates:
\begin{prop}
\label{prop:estphi}
Let $u(x,y)\in H^4(\RR^2)\cap W^{2,2\pm}$ and $(x+iy)^3 u(x,y)\in  W^{2,2\pm}$. Let 
$k$ satisfy the inequality (\ref{eq:invbar7}).

Then
\begin{enumerate}
\item $\phi\in H^4\cap W^{2,2\pm}$
\item 
\beq
\label{eq:estpsi1}
\phi=\frac{c_2}{z^2}+O\left(\frac{1}{z^3}  \right), \ \ |z|\rightarrow\infty,
\eeq
\beq
\label{eq:estpsi1}
\partial_z\phi=O\left(\frac{1}{z^3}  \right),  \ \ |z|\rightarrow\infty. 
\eeq
\end{enumerate}
\end{prop}

The proof consists of a series of steps.
\begin{enumerate}
\item We want to view one of the functions $\phi$ in the
quadratic term of equation (\ref{eq:hopf4}) as a known function ($\phi$) and the other one as the unknown 
($\Phi$). Therefore equation (\ref{eq:hopf4}) can be rewritten as a linear equation for $\Phi$ 
in two different ways:
\beq
\label{eq:hopf6.1}
\partial_{\bar z} \Phi +\frac{1}{\bar k -k}   \phi (\partial_z-\partial_{\bar z})\Phi=
-\frac{1}{\bar k -k} (\partial_z-\partial_{\bar z}) u,
\eeq
and
\beq
\label{eq:hopf6.2}
\partial_{\bar z} \Phi +\frac{1}{2(\bar k -k)} (\partial_z-\partial_{\bar z}) 
[\phi\cdot\Phi]=
-\frac{1}{\bar k -k} (\partial_z-\partial_{\bar z}) u.
\eeq

Function $\Phi$ can be obtained by solving the integral equation
\beq
\label{eq:hopf6.3}
\Phi = - \frac{1}{2(\bar k -k)}(\partial_{\bar z}^{-1}\partial_z-1)[\phi\cdot 
\Phi]-\frac{1}{\bar k -k} (\partial_{\bar z}^{-1}\partial_z-1) u,
\eeq
in all spaces $W^{2,2+\epsilon}$, where $|\epsilon|$ is sufficiently small. Now the iteration process 
converges due to estimates (\ref{eq:invbar8.3}), (\ref{eq:belt2}); therefore $\Phi\in W^{2,2\pm}$, 
and the first statement  of Proposition~\ref{prop:estphi} is proven.

\item Consider the auxiliary {\bf homogeneous} equation:
\beq
\label{eq:beltrami1.1}
\partial_{\bar z} w +\frac{1}{\bar k -k}   \phi (\partial_z-\partial_{\bar z})w=0,
\eeq
which is equivalent to the Beltrami equation:
\beq
\label{eq:beltrami1.2}
\partial_{\bar z} w -q(z,\bar z) \partial_z w=0,
\eeq
where
\beq
\label{eq:beltrami1.3}
q(z,\bar z)=\frac{\frac{\phi(z,\bar z)}{k-\bar k}}
{1+\frac{\phi(z,\bar z)}{k-\bar k}},
\eeq
\beq
\label{eq:beltrami1.4}
|q(z,\bar z)|\le C, \ \ \ \ \ \|q(z,\bar z)\|_{H^l}\le\frac{C}{\alpha_l}, \ \ C<1.
\eeq

To continue, we need the following lemma from \cite{Vekua}:

\begin{lemma}
\label{lem:belt2}
Let $q(z)$ be a measurable bounded function such that
\begin{enumerate}
\item $|q(z)|\le C<1$ for all $z\in\CC$.
\item $q(z)\in L^p$ for some $p<1$.
\end{enumerate}
Then the Beltrami equation (\ref{eq:beltrami1.2}) has an unique (up to a constant) 
solution $w(z)$ such that
\begin{enumerate}
\item $w=z+O(1)$ as $z\rightarrow\infty$
\item $z\rightarrow w(z,\bar z)$ defines a one-to-one continuous map $\overline{\CC}\rightarrow\overline{\CC}$. 
\end{enumerate}
\end{lemma}

We know that $\phi\in H^4(\RR^2)\cap W^{2,2\pm}(\RR^2)$, therefore we can improve the estimate (a) of
Lemma~\ref{lem:belt2}.

\begin{lemma}
Let $q$ be defined by (\ref{eq:beltrami1.3}) in terms of $\phi\in H^4(\RR^2)\cap W^{2,2\pm}(\RR^2)$
satisfying (\ref{eq:invbar8.3}), with $C<1/2$. Then 
\begin{enumerate}
\item $w(z) = z+o(1)$.
\item All functions 
$$
\frac{\partial w}{\partial z}, \ \ \frac{\partial\bar w}{\partial z}, \ \ 
\frac{\partial w}{\partial\bar z}, \ \ \frac{\partial\bar w}{\partial\bar z}, \ 
\frac{\partial z}{\partial w}, \ \ \frac{\partial\bar z}{\partial w}, \ \ 
\frac{\partial z}{\partial\bar w}, \ \ \frac{\partial\bar z}{\partial\bar w}, \ 
$$
$$
\frac{\partial^2 w}{\partial z^2}, \ \ \frac{\partial^2 w}{\partial\bar z^2}, \ \
\frac{\partial^2 w}{\partial z \partial\bar z},\ \
\frac{\partial^2\bar w}{\partial z^2}, \ \ \frac{\partial^2\bar w}{\partial\bar z^2}, \ \
\frac{\partial^2\bar w}{\partial z \partial\bar z},
$$
$$
\frac{\partial^2 z}{\partial w^2}, \ \ \frac{\partial^2 z}{\partial\bar w^2}, \ \
\frac{\partial^2 z}{\partial w \partial\bar w},\ \
\frac{\partial^2\bar z}{\partial w^2}, \ \ \frac{\partial^2\bar z}{\partial\bar w^2}, \ \
\frac{\partial^2\bar z}{\partial w \partial\bar w},
$$
are continuous and bounded. 
\item $Jac((w,\bar w),(z,\bar z))$ and  $Jac((z,\bar z),(w,\bar w))$ are continuous and bounded 
functions.
\end{enumerate}
\end{lemma}
The proof follows from the the following explicit formula 
\beq
\label{eq:beltrami2.2}
w=z+\partial^{-1}_{\bar z}f, \ \ f=\left[\sum\limits_{k=0}^{\infty} 
(q\partial_z\partial^{-1}_{\bar z})^k\right] q
\eeq
The convergence of the series in $W^{2,2\pm}$ follows from arguments similar to those in 
Lemma~\ref{lem:belt1} of the Appendix. In $H^4$ we have the estimate 
\beq
\label{eq:beltrami2.2.1}
\|(q\partial_z\partial^{-1}_{\bar z})\| \le  \alpha_4 \|q(z,\bar z)\|_{H^4} \le C < \frac{1}{2}.
\eeq
Therefore 
\begin{align}
\label{eq:beltrami2.2.2}
&\|f\|_{H^4} \le 2 \|q(z,\bar z)\|_{H^4} \le \frac{2C}{\alpha_4},  \nonumber \\
&\|\partial_z\partial^{-1}_{\bar z}f\|_{H^4} \le 2 \|q(z,\bar z)\|_{H^4} \le \frac{2C}{\alpha_4}.  \ 
\end{align}

From Lemma~\ref{lem:l7}, part 2 we also have 
\beq
\label{eq:beltrami2.2.3}
|f|  \le 2C < 1 ,  \ \ |\partial_z\partial^{-1}_{\bar z}f|  \le 2C <1.  
\eeq

Since $w=z+\partial^{-1}_{\bar z}f$, where $f\in H^4(\RR^2)\cap W^{2,2\pm}(\RR^2)$, 
then $\hat f\in L^q(\RR^2)\cap L^{1}(\RR^2)$ for some $q>2$. From the H\"older inequality 
it follows that $\widehat {\partial^{-1}_{\bar z}  f}\in L^{1}(\RR^2)$, 
and $\partial^{-1}_{\bar z} f = o(1)$ for $z\rightarrow\infty$. 

Since 
\beq
\label{eq:beltrami2.2.4}
\partial_z w =1 + \partial^{-1}_{\bar z} f, \ \ \partial_{\bar z} w = q \partial_z w,
\eeq
one immediately sees that
\beq
\label{eq:beltrami2.2.5}
Jac((w,\bar w),(z,\bar z)) = ( \partial_z w \cdot \overline{\partial_z w} )(1-q\bar q) \ge 
(1- 2C)^2 (1-C^2)>0.
\eeq
\qed

It is natural to use $w$ as a new coordinate. It is easy to check that 
\beq
\label{eq:beltrami2.3}
[\partial_{\bar z}-q\partial_z] = [1-q\bar q]
\overline {\left( \frac{\partial w}{\partial z} \right)}\partial_{\overline{w}}.
\eeq
Equation (\ref{eq:hopf6.1}) is equivalent to:
\beq
\label{eq:hopf7}
\partial_{\overline{w}} \Phi =  {\mathcal U}, \ \ \mbox{where} \ \ 
{\mathcal U}=-[1-q\bar q]^{-1}
\overline {\left( \frac{\partial w}{\partial z} \right)}^{-1} 
\left( 1+\frac{\phi(z,\bar z)}{k-\bar k}\right)^{-1} u_x.
\eeq

Let ${\mathcal U}(w,\bar w)={\mathcal U}_+(w,\bar w)+{\mathcal U}_-(w,\bar w)$, where 
${\mathcal U}_+(w,\bar w)$ has support in the ball $|z|\le 2$ and 
${\mathcal U}_-(w,\bar w)$ has support outside the ball $|z|\le 1$.  
Then 

\beq
\label{eq:hopf8}
\Phi = \Phi_+ + \Phi_-, \ \  \Phi_+= \partial_{\overline{w}}^{-1} {\mathcal U}_+, 
\ \   \Phi_-= \partial_{\overline{w}}^{-1} {\mathcal U}_-.
\eeq

The function $\Phi_+$ is holomorphic in $1/w$ outside the ball  $|z|\le 2$, and 
\beq
\label{eq:hopf9}
\Phi_-= \frac{d_1}{w}+  \frac{d_2}{w^2}+  \frac{d_3}{w^3}+  \frac{1}{w^3}
\partial_{\overline{w}}^{-1} (w^3 {\mathcal U}_-).
\eeq
We know that $\Phi\in W^{2,2\pm}$, therefore, for large $|z|$, we have 
\beq
\label{eq:hopf10}
\Phi= \frac{\tilde d_2}{w^2}+  \frac{\tilde d_3}{w^3}+  \frac{1}{w^3}
\partial_{\overline{w}}^{-1} (w^3 {\mathcal U}_-),
\eeq
\beq
\label{eq:hopf11}
\partial_{w}\Phi= O\left(\frac{1}{w^3}\right), \ \ \partial_{\bar w}\Phi= O\left(\frac{1}{w^3}\right)
\eeq

Here we used, that $(x+iy)^3 u(x,y)\in  W^{2,2\pm}$, therefore $w^3 {\mathcal U}_-\in  W^{1,2\pm}$,
$\partial_{\overline{w}}^{-1} (w^3 {\mathcal U}_-)$ is continuous, bounded and decays at infinity,
the functions $w^3 {\mathcal U}_-\in  W^{1,2\pm}$ and 
$\partial_w\partial_{\overline{w}}^{-1} (w^3 {\mathcal U}_-)$ are continuous and bounded (see 
Lemma~\ref{lem:vek} of the Appendix).

Taking into account that $\partial_{z}w$,  $\partial_{\bar z}w$ are bounded 
continuous functions, we finish the proof of Proposition~\ref{prop:estphi}.
\end{enumerate}\qed

\section{Solution of the linearized Hopf equation}
\label{sec:s8}
Now we are ready to construct the function $\Lambda_2(x,y,k)$.
In the complex coordinates, equation (\ref{eq:hopf2}) has the following form
\beq
\label{eq:hopf12}
\partial_{\bar z}\Lambda_2+ \frac{1}{\bar k -k}(\partial_{z}-\partial_{\bar z})(\phi\Lambda_2)=0,
\eeq
and the boundary condition (\ref{eq:asympt2.2}) is equivalent to 
\beq
\label{eq:asympt4}
\Lambda_2(x,y,k)= (k-\bar k)z+ o(1) \ \ \mbox{as} \ \  z\rightarrow\infty.
\eeq

We shall use the following simple 
\begin{lemma}
\label{lem:l6}
Let $\Xi$ be a solution of (\ref{eq:hopf12}), $p(w)$ be an arbitrary {\bf holomorphic} function of $w$, where 
$w$ is the special solution of (\ref{eq:beltrami1.1}) defined in the previous section.  
Then $p(w)\Xi$ also solves  (\ref{eq:hopf12}).
\end{lemma}

Let us now construct a special solution of (\ref{eq:hopf12}) in the form $\Xi=1+\Xi_1$, where  $\Xi_1$ decays 
at infinity. Then we have 
\beq
\label{eq:hopf13}
\partial_{\bar z}\Xi_1+ \frac{1}{\bar k -k}(\partial_{z}-\partial_{\bar z})(\phi\Xi_1)=
-\frac{1}{\bar k -k}(\partial_{z}-\partial_{\bar z})(\phi)
\eeq
\beq
\label{eq:hopf14}
\Xi_1+ \frac{1}{\bar k -k}(\partial_{z}\partial_{\bar z}^{-1}-1)(\phi\Xi_1)=
-\frac{1}{\bar k -k}(\partial_{z}\partial_{\bar z}^{-1}-1)(\phi)
\eeq
\beq
\label{eq:hopf15}
\Xi_1 = - \left[1 + \frac{1}{\bar k -k}(\partial_{z}\partial_{\bar z}^{-1} -1) \phi \right]^{-1}
\frac{1}{\bar k -k}(\partial_{z}\partial_{\bar z}^{-1}-1)(\phi), 
\eeq
and $\Xi_1\in W^{2,2\pm}$; therefore $\Xi_1\rightarrow 0$ as $|z|\rightarrow\infty$,  and 
$\partial_z\Xi_1$ is a bounded function. Equation (\ref{eq:hopf14}) is equivalent to:
\begin{align}
\label{eq:hopf16}
&\Xi_1= \frac{1}{k -\bar k}(\partial_{z}\partial_{\bar z}^{-1}-1)(\phi\Xi_1)+
\frac{1}{k -\bar k}(\partial_{z}\partial_{\bar z}^{-1}-1)(\phi)=\nonumber \\
&=\frac{1}{k -\bar k}\left[-\phi -\phi\Xi_1 + \partial_{\bar z}^{-1} \left( \phi_z + \phi_z\Xi_1 + 
\phi (\Xi_1)_z \right) \right]= \nonumber \\
&=O\left(\frac{1}{z^2}\right) +  \partial_{\bar z}^{-1}O\left(\frac{1}{z^3}\right)+ 
\frac{1}{k -\bar k} \partial_{\bar z}^{-1} \left(\frac{c_2}{z^2}  (\Xi_1)_z \right)=\\
&=O\left(\frac{1}{z^2}\right)+ \frac{d'}{z}+ O\left(\frac{1}{z^{2-\epsilon}}\right)+\frac{d''}{z}+
+\frac{d'''}{z^2}+\frac{c_2}{z^2} \partial_{\bar z}^{-1} (\Xi_1)_z= \nonumber\\
&= \frac{d}{z}+ O\left(\frac{1}{z^{2-\epsilon}}\right),\ \ z\rightarrow\infty,  \nonumber
\end{align}
where $\epsilon>0$ is an arbitrary positive constant, and $d$, $d'$, $d''$,  $d'''$ are some momenta. Since
$\Xi_1\in W^{2,2\pm}$, then $d=0$ and
\beq
\label{eq:hopf17}
\Xi = 1 +  O\left(\frac{1}{z^{2-\epsilon}}\right).
\eeq
Using Lemma~\ref{lem:l6}, we finally obtain:
\beq
\label{eq:hopf18}
\Lambda_2 = 2 \Im k\cdot \Xi\cdot w
\eeq

At last, from equations (\ref{eq:invbar10}), (\ref{eq:invbar11}), (\ref{eq:hopf12}) we obtain the asymptotics 
of $\Lambda_2$ for large $\Im k$:
\begin{align}
\label{eq:hopf19}
\Lambda_2 = &- (x - k y) +\frac{yu}{k} - \frac{xu+\partial_x^{-1}(2yu_y +u) }{k^2} + \\
&+\frac{x\partial_x^{-1}u_y + \partial_x^{-1}(xu_y)+\frac{3}{2} yu^2 +3  \partial_x^{-2}(yu_y)_y}{k^3} 
+O\left(\frac{1}{k^4} \right). \nonumber
\end{align}

\section{From the Hopf equation to the vector fields eigenfunctions}

In the previous section we constructed the function 
$\Lambda_1(x,y,k)=k+\phi(x,y,k)$; i.e. the 
level sets for the function $\Psi_1(x,y,\lambda)$:
\beq
\label{eq:levels3}
\Psi_1(x,y,\Lambda_1(x,y,k))=k.
\eeq
Therefore, to construct $\Psi_1(x,y,\lambda)$, we have to solve with respect to $k$ 
the following equation
\beq
\label{eq:levels4}
\Lambda_1(x,y,k)=\lambda.
\eeq

Let us calculate $\partial_k \Lambda_1(x,y,k)$,  
$\partial_{\bar k} \Lambda_1(x,y,k)$. Differentiating by $k$, $\bar k$ the
principal Hopf equation (\ref{eq:hopf1}) for $\Lambda_1$, we obtain that both 
functions $\partial_k \Lambda_1(x,y,k)$, $\partial_{\bar k} \Lambda_1(x,y,k)$  satisfy 
(\ref{eq:hopf12}) with the boundary conditions:
\beq
\label{eq:asympt5}
\partial_k\Lambda_1(x,y,k)= 1+ o(1), \ \ \partial_{\bar k}
\Lambda_1(x,y,k)= o(1)  \ \ \mbox{as} \ \  z\rightarrow\infty.
\eeq
Therefore
\beq 
\label{eq:hopf20}
\partial_k\Lambda_1(x,y,k)= \Xi=1+\Xi_1 , \ \ \partial_{\bar k}
\Lambda_1(x,y,k)\equiv 0.
\eeq
In particular, $\Lambda_1(x,y,k)$ is holomorphic in $k$.
To calculate $\partial_{\bar k} \Lambda_2(x,y,k)$ we can differentiate by 
$\bar k$ equation (\ref{eq:hopf2.2}). Taking into account that  
$\Lambda_1(x,y,k)$ is holomorphic in $k$, we obtain
\beq
\label{eq:hopf21}
(\partial_{\bar k} \Lambda_2)_y+(\Lambda_1\partial_{\bar k} \Lambda_2)_x=0,
\eeq
with the boundary condition
\beq 
\label{eq:hopf22}
\partial_{\bar k} \Lambda_2(x,y,k)= o(1) \ \ \mbox{as} \ \  z\rightarrow\infty;
\eeq
therefore 
\beq 
\label{eq:hopf23}
\partial_{\bar k} \Lambda_2(x,y,k)\equiv0.
\eeq
Let $u(x,y)$ satisfy (\ref{eq:invbar7}) with $C<1/2$, $l=4$.
From (\ref{eq:hopf15}) and (\ref{eq:invbar8.3}) we immediately obtain that 
\beq
\label{eq:invbar13}
\|\Xi_1\|_{H^4} \le \frac{2C}{\alpha_4},  \ \ \max|\Xi_1|\le 2C,  
\ \ C<\frac{1}{2}.
\eeq

Let us fix a point $(x,y)$ and consider the map 
\beq
\label{eq:map1}
k\rightarrow \lambda=\Lambda_1(x,y,k)
\eeq
from the semi-plane
\beq
\label{eq:map2}
\Im k\ge D=\sqrt{\frac{2\alpha_4\|u\|_{H^4}}{C}} 
\eeq
to the complex plane. We see that, for a pair of points $k_1$, $k_2$,
\beq
\label{eq:map3}
\Lambda_1(x,y,k_1)- \Lambda_1(x,y,k_2)=k_1-k_2 + \phi(x,y,k_1)-\phi(x,y,k_2).
\eeq
But 
\beq
\label{eq:map4}
|\phi(x,y,k_1)-\phi(x,y,k_2)|\le |k_1-k_2| \max\limits_{k\in[k_1,k_2]} 
|\Xi_1(x,y,k)|\le  2C |k_1-k_2|,
\eeq
where $[k_1,k_2]$ denotes the segment connecting the points $k_1$ and $k_2$.

Therefore 
\beq
\label{eq:map5}
|\Lambda_1(x,y,k_1)- \Lambda_1(x,y,k_2)|\ge |k_1-k_2|(1-2C),
\eeq
and different points have different images. 

From (\ref{eq:invbar8.3}) it follows that the image completely covers the 
semi-plane 
\beq
\label{eq:map6}
\Im\lambda \ge  \sqrt{\frac{2\alpha_4\|u\|_{H^4}}{C}} \cdot(1+C), \ \ C<\frac{1}{2}.
\eeq

To check that a point $\lambda_0$, belongs to this image it is sufficient to consider 
the image of the square with the boundaries
\beq
\Im k = D, \ \ \Im k= \frac{2 \Im\lambda_0 }{1-C}, \ \ \Re k = \Re \lambda_0 -  
\frac{2 \Im\lambda_0 }{1-C}, \ \  \Re k =  \Re \lambda_0 + 
\frac{2 \Im\lambda_0 }{1-C}.
\eeq

An easy estimate shows that $\lambda_0$ lies inside the region surrounded by the image of the
above boundary, therefore $\lambda_0$ has a preimage, and, as it was shown above, this 
preimage is unique. 
 
Therefore, on the semi-plane defined by (\ref{eq:map6}), the functions
$\Psi_1(x,y,\lambda)$, $\Psi_2(x,y,\lambda)$  are well-defined and 
holomorphic in $\lambda$.

We end this paper rederiving the well-known formal asymptotics of 
$\Psi_1$ and $\Psi_2$ for large $|\Im\lambda|$. 

Combining equations (\ref{eq:ham7}) and (\ref{eq:invbar11}) we obtain the asymptotic formulas:
\beq
\label{eq:asympt6}
\lambda = \Lambda_1=k-\frac{u}{k}+\frac{\partial_y\partial_x^{-1} u}{k^2}-
\frac{\partial_y^2\partial_x^{-2} u + u^2/2   }{k^3} 
+O\left(\frac{1}{k^4} \right), \ \ 
|\Im k|\rightarrow\infty.
\eeq
Then the inversion with respect to $k$ yields
\beq
\label{eq:asympt7}
\Psi_1(x,y,\lambda)=k=\lambda +\frac{u}{\lambda}-\frac{\partial_x^{-1} u_y}{\lambda^2}+
\frac{\partial_x^{-2} u_{yy} - u^2/2   }{\lambda^3} 
+O\left(\frac{1}{\lambda^4} \right), \ \ |\Im\lambda|\rightarrow\infty.
\eeq
In addition, combining (\ref{eq:psi2def}), (\ref{eq:asympt7}) and (\ref{eq:hopf19}), one
obtains
\beq
\label{eq:asympt8}
\Psi_2(x,y,\lambda)=x-\lambda y -\frac{yu}{\lambda}+ \frac{\partial_x^{-1}(y u)_y}{\lambda^2}+
\frac{\frac{yu^2}{2} - \partial_x^{-2} (yu)_{yy} }{\lambda^3} 
+O\left(\frac{1}{\lambda^4} \right), \ \ |\Im\lambda|\rightarrow\infty.
\eeq

\section{Appendix. Some basic fact from functional analysis}

Let us recall some basic facts about Banach spaces $L^p(\RR^2)$. 

\begin{enumerate}
\item H\"older inequality (see \cite{RS2}). Let
\beq
\label{eq:fourier6}
1\le p,q,r \le \infty, \ \ \frac{1}{p}+\frac{1}{q}=\frac{1}{r},
\eeq
$f\in L^p$, $g\in L^q$.
Then
\beq
\label{eq:fourier7}
f\cdot g\in L^r, \ \ \mbox{and} \ \ \left\| f\cdot g\right\|_{L^r}\le
\left\| f\right\|_{L^p}\left\| g\right\|_{L^q}.
\eeq
\item Young's inequality (see \cite{RS2}). Let,
\beq
\label{eq:fourier8}
1\le p,q,r \le \infty, \ \ \frac{1}{p}+\frac{1}{q}=1+\frac{1}{r},
\eeq
$f\in L^p$, $g\in L^q$.
Then
\beq
\label{eq:fourier9}
f * g\in L^r, \ \ \mbox{and} \ \ \left\| f*g\right\|_{L^r}\le
\left\| f\right\|_{L^p}\left\| g\right\|_{L^q}.
\eeq
\item Hausdorff-Young inequality  (see \cite{RS2},  \cite{Katz}).

It is well-known that the Fourier transform (\ref{eq:fourier1}) is a unitary
map $L^2(\RR^2)\rightarrow L^2(\RR^2)$. For Banach spaces the situation is
more delicate:
Let  
\beq
\label{eq:fourier10}
1\le p \le 2, \ \ \frac{1}{p}+\frac{1}{q}=1
\eeq
Then the Fourier transform is a well-defined map 
$L^p(\RR^2)\rightarrow L^q(\RR^2)$ and
\beq
\label{eq:fourier11}
\left\| \hat f \right\|_{L^q}\le (2\pi)^{(1-2/p)}  \left\| f\right\|_{L^p},
\eeq
but this map is {\bf not invertible}. If $f(z)\in L^{2}\cap L^{p}$, then  
its Fourier image belongs to $L^2$, but this estimate can not be improved.
\end{enumerate}

We need the following properties of the Sobolev spaces 
(see the books \cite{Adams}, \cite{BenzSer}, \cite{Wells} )

\begin{lemma}
\label{lem:l7}
Let $l\ge 2$. Then 
\begin{enumerate}
\item $H^l(\RR^2)\subset C^{l-2}(\RR^2)$. Moreover, there exists a constant $\beta_l$ such that, for 
$f(x,y)\in H^l(\RR^2)$,
\beq
\label{eq:sobolev3}
\max\limits_{(x,y)\in\RR^2} |f(x,y)|\le \beta_l \|f\|_{H^l}.
\eeq
There are similar estimates also for the first $l-2$ derivatives. 
\item $H^l(\RR^2)$ is a Banach algebra, i.e. is closed with respect to the multiplication 
and there exists a constant $\alpha_l$ such that, for any $f,g \in H^l(\RR^2)$,
\beq
\label{eq:sobolev4}
\|fg\|_{H^l} \le\alpha_l \|f\|_{H^l} \|g\|_{H^l}, 
\eeq
$\beta_l\le\alpha_l$.
\item Let $f\in H^l(\RR^2)$. Then the multiplication operator 
$$
h\rightarrow f\cdot h 
$$
is a bounded operator on all spaces $W^{l',p}$, $l'\le l-2$.
\item The operator $\Pi= \partial_{z}\partial_{\bar z}^{-1}$ is well-defined on all spaces 
$W^{l,p}$, $1<p<\infty$, and
\beq
\label{eq:belt3}
\| \partial_{z}\partial_{\bar z}^{-1} f\|_{W^{l,p}} \le \gamma(p) \| f\|_{W^{l,p}}
\eeq
\end{enumerate}
\end{lemma}
Some numerical estimates on the constants $\alpha_l$ were obtained 
in the papers \cite{MP1}, \cite{MP2}. The property (\ref{eq:belt3}) follows immediately from 
Theorem~\ref{th:vekua1}.

\begin{lemma}
\label{lem:vek}
Consider the space $W=W^{2,2-\epsilon}\cap W^{2,2+\epsilon'}$, $0<\epsilon<1$, $\epsilon'>0$,
$\|f\|_{W}= \|f\|_{W^{2,2-\epsilon}}+\|f\|_{ W^{2,2+\epsilon'}}$.

\begin{enumerate}
\item Let $f\in W$, then $f$, $f_z$, $f_{\bar z}$ are H\"older functions for some $\alpha>0$
(and, as a corollary, continuous), bounded and $f(z)=o(1)$ as $z\rightarrow\infty$.
\item $W$ is a Banach algebra, i.e. if $f,g\in W$, then $f\cdot g\in W$ and 
$\|f\cdot g \|_W\le C(\epsilon,\epsilon') \|f\|_W \|g \|_W $.
\end{enumerate}
\end{lemma}

This Lemma follows immediately from the Theorem~1.21 from Vekua's book 
\cite{Vekua} stating that:

if $f\in L^p(\RR^2)\cap\L^{p'}(\RR^2)$ with $p>2$, $1<p'<2$, then for 
$g=\partial_{\bar z}^{-1}f$ we have the inequalities:
$$
|g(z)| \le M_{p,p'} (\|f\|_{L_p}+ \|f\|_{L_p'}), \ \ z\in\CC,
$$
$$
|g(z_1)-g(z_2) | \le M_{p,p'} (\|f\|_{L_p}+ \|f\|_{L_p'})
|z_1-z_2|^{\frac{p-2}{p}}.
$$ 

\begin{lemma}
\label{lem:belt1}
Let $q(z,\bar z)$ be a two times continuously differentiable function, such that 
all functions $q$, $q_{z}$,  $q_{\bar z}$,  $q_{zz}$,  $q_{z\bar z}$,
$q_{\bar z\bar z}$ are bounded, and, in addition, $|q|\le C<1$. 

Then there exists an $\epsilon>0$ such that the operator 
\beq
\label{eq:belt4}
B=(1-q(z,\bar z) \partial_{z}\partial_{\bar z}^{-1})  
\eeq
is invertible in all spaces $W^{2,2+\epsilon_1}$, $-\epsilon<\epsilon_1<\epsilon$, and 
the inverse operator $B^{-1}$ is uniformly bounded in $\epsilon_1$.
\end{lemma}

The proof is based on the following trick. Consider the space $W^{2,p}(\RR^2)$. In addition 
to the norm, $\| \|_{W^{2,p}}$, it is convenient to introduce a family of norms 
 $\| \|_{W^{2,p}_{\mu}}$, which are all equivalent for a fixed $k$:
\begin{align}
\label{eq:sobolev5}
& \|f\|_{W^{l,p}_{\mu}}  = & \left[
\iint |f|^p dx dy + 
\mu^{p} \iint (|f_z|^p+ |f_{\bar z}|^p)   dx dy +  \right. & \\
& &  + \left. \mu^{2p} \iint (|f_{zz}|^p  + 2 |f_{z\bar z}|^p + 
|f_{\bar z\bar z}|^p) dx dy  \right]^{1/p}, \ \ & \mu>0. \nonumber
\end{align}

Let $q(z,\bar z)$ be a two times continuously differentiable function, and 
all functions $q$, $q_{z}$,  $q_{\bar z}$,  $q_{zz}$,  $q_{z\bar z}$,  
$q_{\bar z\bar z}$ be bounded, $|q|\le C$. Then the multiplication operator 
\beq
\label{eq:belt4}
f \rightarrow q\cdot f
\eeq
is a well-defined bounded operator on $W^{2,p}$. Moreover, it is easy to check that 
\begin{align}
\label{eq:belt5}
\|q\cdot f\|_{W^{l,p}_{\mu}} \le \left[ \max\limits_{z\in\CC} |q| + 2\mu 
\left(\max\limits_{z\in\CC} |q_z|+ \max\limits_{z\in\CC} |q_{\bar z}|  \right) +
\right. \\
+\left. \mu^2 
\left(\max\limits_{z\in\CC} |q_{zz}|+ \max\limits_{z\in\CC} |q_{\bar z\bar z}|+ 2
\max\limits_{z\in\CC} |q_{z\bar z|}  \right) \right]\|f\|_{W^{l,p}_{\mu}}
\end{align}

We shall use also the following simple
 
\begin{lemma}
Let $z\rightarrow w(z,\bar z)$ be a one-to-one map 
$\overline{\CC}\rightarrow\overline{\CC}$ such that
\begin{enumerate}
\item Both functions $w(z,\bar z)$ and $z(w,\bar w)$ are two times continuously 
differentiable.
\item All functions 
$$
\frac{\partial w}{\partial z}, \ \ \frac{\partial\bar w}{\partial z}, \ \ 
\frac{\partial w}{\partial\bar z}, \ \ \frac{\partial\bar w}{\partial\bar z}, \ 
\frac{\partial z}{\partial w}, \ \ \frac{\partial\bar z}{\partial w}, \ \ 
\frac{\partial z}{\partial\bar w}, \ \ \frac{\partial\bar z}{\partial\bar w}, \ 
$$
$$
\frac{\partial^2 w}{\partial z^2}, \ \ \frac{\partial^2 w}{\partial\bar z^2}, \ \
\frac{\partial^2 w}{\partial z \partial\bar z},\ \
\frac{\partial^2\bar w}{\partial z^2}, \ \ \frac{\partial^2\bar w}{\partial\bar z^2}, \ \
\frac{\partial^2\bar w}{\partial z \partial\bar z},
$$
$$
\frac{\partial^2 z}{\partial w^2}, \ \ \frac{\partial^2 z}{\partial\bar w^2}, \ \
\frac{\partial^2 z}{\partial w \partial\bar w},\ \
\frac{\partial^2\bar z}{\partial w^2}, \ \ \frac{\partial^2\bar z}{\partial\bar w^2}, \ \
\frac{\partial^2\bar z}{\partial w \partial\bar w},
$$
are bounded. 
\item $Jac((w,\bar w),(z,\bar z))$ and  $Jac((z,\bar z),(w,\bar w))$ are bounded
functions.
\end{enumerate}
Then the corresponding map $W^{2,p}(\CC)\rightarrow W^{2,p}(\CC)$: 
$f(w,\bar w)\rightarrow \tilde f(z,\bar z) = f(w(z,\bar z),\bar w(z,\bar z))$ 
is well-defined and bounded in both directions.
\end{lemma}

\thanks{Acknowledgments. We would like to thank C.Morosi for explaining us some results on Sobolev 
spaces, and M.Pavlov for pointing to us some important references.}

\end{document}